\documentclass[11pt, a4paper]{article}
\usepackage{blindtext}
\usepackage[a4paper, total={6in, 8in}]{geometry}
\usepackage[utf8]{inputenc}
\usepackage{enumitem}
\usepackage{amssymb}
\usepackage{graphicx}
\usepackage{physics}
\usepackage{mathtools}
\usepackage{subcaption}

\usepackage[font=small,labelfont=bf]{caption} 
\usepackage{amsmath}
\usepackage[affil-it]{authblk}
 
\title{  Diagrammatic perturbation theory for Stochastic nonlinear oscillators}
\author{Akshay Pal, Jayanta Kumar Bhattacharjee
  \thanks{Electronic address: \texttt{akshayphysics1804@gmail.com,tpjkb@iacs.res.in}}}
\affil{School of Physical Sciences\\Indian Association for the Cultivation of Science,Jadavpur,Kolkata,India}
\date{} 
\begin{document}
\maketitle
\begin{abstract}
We consider the stochastically driven one dimensional nonlinear oscillator
$\ddot{x}+2\Gamma\dot{x}+\omega^2_0 x+\lambda x^3 = f(t)$ where f(t) is a Gaussian noise which, for the bulk of the work, is delta correlated (white noise). We construct the linear response function in frequency space in a systematic Feynman diagram-based perturbation theory. As in other areas of physics, this expansion is characterized by the number of loops in the diagram. This allows us to show that the damping coefficient acquires a correction at $O(\lambda^2)$ which is the two loop order. More importantly, it leads to the numerically small but conceptually interesting finding that the response is a function of the
frequency at which a stochastic system is probed. The method is easily generalizable to different kinds of nonlinearity and replacing the nonlinear term in the above equation by $\mu x^2$ , we can discuss the issue of noise driven escape from a potential well. If we add a periodic forcing to the cubic nonlinearity situation, then we find that the response function can have a contribution jointly proportional to the strength of the noise and the amplitude of the periodic drive. To treat the stochastic Kapitza problem in perturbation theory we find that it is necessary to have a coloured noise.
\end{abstract}
\section{Introduction} 
Stochastic nonlinear oscillators have been investigated extensively [1-10] from the perspective  of  dynamical systems [11-12] over the last two decades. The focus, very often, has been on an added stochasticity on the natural frequency. For the study of dynamics, the primary goal is to see how the different periodic orbits (fixed points,centres or limit cycles )are affected in the presence of noise and what effect does stochasticity have on the possible bifurcations [13-15]. For investigations in statistical physics , the issues have been decay of metastable states, oscillations between multiple stable states etc. Our goal in this work is to systemize the perturbation theory of nonlinear stochastic oscillators along the line of diagrammatic perturbation theory which has been a very useful tool in many areas of physics. In fact a very early attempt in this direction was made by Kraichnan [16] for Langevin dynamics and was followed by a detailed application by Wyld [17] to turbulence in randomly stirred fluids described by the Navier-Stokes equation. In this work , we systematically introduce the Feynman diagram expansion technique to stochastically perturbed dynamical systems.
Our approach in some sense is complementary to that of Belousov [18] who have recently developed a Volterra series approach for stochastic nonlinear dynamics which is particularly effective for the bistable oscillators.
The outcome of this exercise is the unveiling of small effects which had not previously been considered. The primary point is that the linear response function which so far has been characterized by a shift in the natural frequency is now more complex with a change in the natural frequency and the damping coefficient as well . What is more important conceptually is that the two constituents of the response function – the natural frequency and the damping coefficient – are now  weakly dependent on the frequency at which the response is being studied. The situation is analogous to the small but detectable effects in second order phase transitions- the anomalous dimension [19] and the weakly frequency dependent transport coefficients [20-23].\\
We explain our goal by focussing on the stochastically driven damped nonlinear oscillator with a cubic nonlinearity. We will call this the stochastic Duffing oscillator. The equation of motion is taken to be 
\begin{equation}
\label{eq:1}
    \ddot{x}+2\Gamma\dot{x}+\omega_0^2x+\lambda x^3=f(t)
\end{equation}
The variable $x(t)$ is the displacement of the nonlinear oscillator, the constant $\Gamma$ is the damping coefficient and $f(t)$ is the random force. The random force is taken to have a Gaussian distribution with the probability $P[f(t)]\varpropto exp[-\int_0^\infty f(t)^2 dt/2D ]$  for the realization of a given $f(t)$. This distribution implies that all the odd correlations of $f(t)$ vanish and the even correlations have the Gaussian factoring given by
\begin{equation}
    <f(t_1)f(t_2)...f(t_{2n})>=\sum_{all\, pairings}<f(t_1)f(t_2)>....<f(t_{2n-1})f(t_{2n})>
\end{equation}
with  ( for the white noise which is the most popular and technically easiest )
\begin{equation}
\label{eq:3}
    <f(t)f(t')>=2D\delta(t-t')
\end{equation}
The angular brackets denote an averaging over the noise distribution.The only time we depart from this prescription will be in Sec V.The perturbation theory proceeds by expanding 
\begin{equation}
\label{eq:4}
    x(t)=x_0(t)+\lambda x_{1}(t)+\lambda^2 x_2(t)+....
\end{equation}
Inserting  this expansion in Eq.\ref{eq:1} and equating the coefficients of $\lambda^n$ on either side of the equation for n=1,2,3,4...,we get
\begin{subequations}
\begin{equation}
\label{eq:5.a}
    \ddot{x_0}+2\Gamma\dot{x_0}+\omega^2_0x_0=f(t)
\end{equation}
\begin{equation}
\label{eq:5.b}
    \ddot{x_1}+2\Gamma\dot{x_1}+\omega^2_0x_1=-x^3_0
\end{equation}
\begin{equation}
\label{eq:5.c}
     \ddot{x_2}+2\Gamma\dot{x_2}+\omega^2_0x_2=-3{x_0}^2x_1
\end{equation}
\end{subequations}
It is easier to work in frequency space using the Fourier transform:
\begin{subequations}
\begin{equation}
    x(\omega)=\int_0^\infty x(t)e^{-i\omega t}dt
\end{equation}
and the inverse
\begin{equation}
    x(t)=\int_{\infty}^{\infty}x(\omega)e^{i\omega t}\frac{d \omega}{2\pi}
\end{equation}
\end{subequations}
In Fourier space, Eq.\ref{eq:1} becomes
\begin{equation}
    (\omega^2_0-\omega^2+2i\Gamma \omega)x(\omega)=f(\omega)-\lambda\int\frac{d\omega_1}{2\pi}\frac{d\omega_2}{2\pi}x(\omega_1)x(\omega_2)x(\omega-\omega_1-\omega_2)
\end{equation}
We  will follow the convention that  when not specified the frequency integration will cover the entire domain $-\infty\leq\omega\leq\infty$.
The linear response function $R(\tau)$ of the system is defined by the relation
\begin{equation}
    x(t)=\int dt' R(t-t')f(t')
\end{equation}
Clearly, for causality,$R(t-t')$ can be non zero only if $t>t'$ and hence
\begin{equation}
     x(t)=\int_{0}^{\infty} d\tau R(\tau)f(t-\tau)
\end{equation}
with $R(\tau)=0, \tau<0$.The Fourier transform of the response function is given by
\begin{equation}
    R(\omega)=\int_0^{\infty} R(t)e^{-i\omega t}dt
\end{equation}
Clearly $R(\omega)$ has no poles in the lower half plane. For calculational purposes it is convenient to define $R(\omega)$ by the functional derivative
\begin{equation}
    R(\omega)=\Big< \frac{\delta x(\omega)}{\delta f(\omega)} \Big>
\end{equation}
The angular bracket denotes an average over the noise performed in Fourier space. The structure of the equation of motion Eq.\ref{eq:1} suggests an alternative prescription for the response function 
\begin{equation}
\label{eq:12}
    R(\omega)=\frac{< x(\omega)f(\omega^{'}) >}{< f(\omega)f(\omega^{'}) >}
\end{equation}
In Fourier space, Eq.\ref{eq:4} becomes
\begin{equation}
    x(\omega)=x_0(\omega)+\lambda x_{1}(\omega)+\lambda^2 x_2(\omega)+....
\end{equation}
Inserting this expansion in Eq.\ref{eq:12} yields   as the expansion
\begin{align}
\label{eq:14}
      R(\omega) &=\frac{1}{< f(\omega)f(\omega^{'}) >}\Big[ < x_0(\omega)f(\omega') > + \lambda < x_1(\omega) f(\omega^{'}) > + \lambda^2 < x_2(\omega) f(\omega^{'}) > + .... \Big] \nonumber\\
      &= R_0(\omega) + \lambda R_1(\omega) + \lambda^2 R_2(\omega) + ...
\end{align}
We now write down the Fourier transforms of (Eq.\ref{eq:5.a})-(Eq.\ref{eq:5.c}) as 

\begin{subequations}
\begin{alignat}{4}
 \Big [ (\omega^2_0-\omega^2)+2i\Gamma \omega \Big] x_0(\omega)&=f(\omega) \label{eq:15.a}\\
 \Big [ (\omega^2_0-\omega^2)+2i\Gamma \omega \Big] x_1(\omega)&=-\int\frac{d\omega_1}{2\pi}\frac{d\omega_2}{2\pi}x_{0}(\omega_1)x_{0}(\omega_2)x_{0}(\omega-\omega_1-\omega_2) \label{eq:15.b}\\
   \Big [ (\omega^2_0-\omega^2)+2i\Gamma \omega \Big] x_2(\omega)&=-3\int\frac{d\omega_1}{2\pi}\frac{d\omega_2}{2\pi}x_{0}(\omega_1)x_{0}(\omega_2)x_{1}(\omega-\omega_1-\omega_2) \label{eq:15.c}
\end{alignat}
\end{subequations}
  We would like to point out that at every order of perturbation theory, the contribution to $R(\omega)$  will be determined by the zeroth order response function 
  \begin{subequations}
  \begin{equation}
  \label{eq:16a}
      R_0(\omega)=\frac{1}{(\omega^2_0-\omega^2)+2i\Gamma \omega}
  \end{equation}
and the zeroth order correlation function $C_{0}(\omega)$ , defined by the relation $C_0(\omega)\delta(\omega-\omega^{'})=< x_0(\omega) x_0(\omega^{'})>$ .Using Eq.\ref{eq:3} and Eq.\ref{eq:15.a}, we  find 
\begin{equation}
\label{eq:16b}
    C_{0}(\omega)=\frac{2D}{(\omega^2_0-\omega^2)^2+4\Gamma^2\omega^2}
\end{equation}
  \end{subequations}
  
It is convenient to write the full response function $R(\omega)$ as
\begin{equation}
\label{eq:17}
    R^{-1}=\Omega^2-\omega^2+2i\omega\Delta
\end{equation}
where $\Omega$ is the dressed or renormalized frequency and $\Delta$ is the corresponding dressed or renormalized damping coefficient. These are the quantities that will be measured as the frequency and as the damping coefficient in any experimental or real life set-up involving the nonlinear oscillator. The renormalized frequency and the damping coefficient will have the perturbation expansion 
\begin{subequations}
\begin{alignat}{4}
 \Omega^2 &=\omega^2_0+\lambda\omega^2_1+\lambda^2\omega^2_2+.....\label{eq:18.a}\\
 \Delta &=\Gamma+\lambda\Gamma_{1}+\lambda^2\Gamma_{2}+..
\end{alignat} \label{eq:18.b}
\end{subequations}
The quantities $\omega^2_{i}$ and $\Gamma_{i}$ for i=1,2,3.....will depend  on the parameters $\omega_0$,$\Gamma$,D  and the frequency $\omega$ at which the system is being probed. Inserting the expansions for $\Omega^2$ and $\Delta$ (Eq.\ref{eq:18.a} and Eq.\ref{eq:18.b}) in Eq.\ref{eq:17}, we arrive at
\begin{align}
\label{eq:19}
    R^{-1}(\omega)&=\omega^2_0-\omega^2+2i\lambda\omega+\lambda(\omega^2_1+2i\omega \Gamma_{1})+\lambda^2(\omega^2_2+2i\omega \Gamma_{2}) \nonumber\\
    &=R^{-1}_0+\lambda(\omega^2_1+2i\omega \Gamma_{1})+\lambda^2(\omega^2_2+2i\omega \Gamma_{2})
\end{align}
Calculating $R^{-1}$ order by order in perturbation theory gives , in general , complex contributions at every order and thus yields the corrections to frequency and damping.\\
In actual practice one calculates $R(\omega)$ in perturbation theory since it is given by a correlation function  Eq.\ref{eq:12} which is straightforward to calculate. We will have to extract information about $R^{-1}(\omega)$ by calculating $R(\omega)$.It should be noted that inverting Eq.\ref{eq:19}, we get
\begin{equation}
\label{eq:20}
    R(\omega)=R_{0}(\omega)-\lambda R_{0}(\omega)^2[\omega^2_1+2i\omega\Gamma_1]+\lambda^2 [R_{0}^3(\omega)(\omega_1^2+2i\Gamma_1\omega
    )^2-R_{0}^2(\omega)(\omega^2_{2}+2i\omega\Gamma_2)]
\end{equation}
The contribution at $O(\lambda)$ is straightforward and from it the lowest order corrections to frequency (real part) and damping (imaginary part) can be directly read off. But at $O(\lambda^2)$ , there are two kinds of contributions. The first of the two $O(\lambda^2)$  terms in Eq.\ref{eq:20} is a product of the two factors of the first along with a response function. It adds on to the first order term as the second term of an emerging geometric series and contributions of this sort are called reducible. The term which cannot be reduced to a product of simpler terms is the irreducible part and from it $\omega^2_2$ and $\lambda_2$  can be read off. This pattern persists at all order of perturbation theory. The conclusion of the above discussion can be summarised as follows:\\
We need to calculate the response function $R(\omega)$ order by order in perturbation theory. At every order we need to identify the irreducible terms – terms which do not factor into products of simpler terms. From these terms we read off the correction to the square of the frequency $\omega^2$ and the correction to the damping $\Gamma$ at that order.\\
The layout of the paper is as follows. In Sec II, we carry out the perturbation theory as described in this section explicitly for the cubic anharmonic oscillator. This is done to second order in the anharmonicity. We will see that the frequency is renormalized at $O(\lambda)$ itself. The damping however picks up a correction term only in the second order in $\lambda$ . The frequency correction at this order is seen to be dependent on the frequency at which the system is being probed. Both these effects would not be obtainable without this formal perturbation theory. In the next section we deal with the nonlinear oscillator which has a periodic driving force in addition to the forces shown in Eq.\ref{eq:1}.The unexpected result here is that there is a frequency shift at $O(\lambda^2)$ which is proportional to the product $\lambda^2A^2D$  where $A$ is the amplitude of the periodic forcing and $D$ quantifies the strength of the stochastic term (see Eq.\ref{eq:3}). The question of escaping from a potential barrier due to the stochastic forcing is treated in Sec.IV and comparing with the barrier crossing probability in statistical mechanics we can find a connection between the parameters $D$ and $\Gamma$ with the temperature $T$. In Sec.V we demonstrate how our formalism leads to the possibility, under special circumstances, of stabilization of an inverted pendulum by stochastic resonance. .We conclude with a brief summary in Sec VI.
\newpage
\section{Diagrammatic perturbation theory}
In this section, we explicitly develop a perturbation theory for the stochastic oscillator with cubic nonlinearity , following the diagrammatic technique which is used in several branches of physics, Our starting point is Eqs \ref{eq:15.a}-\ref{eq:15.c}. From Eq.\ref{eq:15.a} we immediately find $R_0(\omega)$ as given in Eq.\ref{eq:16a}. Our task in this section begins with the calculation of $R_1(\omega)$. For this we use Eqs \ref{eq:14} and \ref{eq:15.b} to obtain
\begin{align}
\label{eq:21}
    R_1(\omega)&=\frac{<x_1(\omega)f(\omega^{'})>}{<f(\omega)f(\omega^{'})>} \nonumber\\
    &=\frac{- R_{0}(\omega)}{<f(\omega)f(\omega^{'})>}\int\frac{d\omega_1}{2\pi}\frac{d\omega_2}{2\pi}<x_0(\omega_1)x_0(\omega_2)x_0(\omega-\omega_1-\omega_2)f(\omega^{'})> \nonumber \\
    &=\frac{-3R_{0}(\omega)}{<f(\omega)f(\omega^{'})>}\int\frac{d\omega_1}{2\pi}\frac{d\omega_2}{2\pi}<x_0(\omega_1)x_0(\omega_2)><x_0(\omega-\omega_1-\omega_2)f(\omega^{'})> \nonumber \\
    &=\frac{-3R_{0}(\omega)}{<f(\omega)f(\omega^{'})>}\int\frac{d\omega_1}{2\pi}\frac{d\omega_2}{2\pi}C_0(\omega_1)\delta(\omega_1+\omega_2)R_0(\omega)<f(\omega)f(\omega^{'})> \nonumber \\
    &=-6DR^2_0(\omega)\int_{-\infty}^{\infty}\frac{d\omega_1}{2\pi}\frac{1}{(\omega^2_0-\omega^2_1)^2+4\Gamma^2\omega^2_1} \nonumber\\
    &=-\frac{3DR^2_0}{2\Gamma \omega^2_{0}}
\end{align}
Comparing with Eq.\ref{eq:20}, we get by equating the real and imaginary parts
\begin{equation}
\label{eq:22}
    \omega^2_1=\frac{3D}{2\gamma\omega^2_{0}} \,\, ,\, \Gamma_1 = 0
\end{equation}
Hence at this order the damping is left unchanged by the nonlinearity and the frequency becomes
\begin{equation}
\label{eq:23}
    \Omega^2=\omega^2_{0}+\frac{3D\lambda^2}{2\Gamma\omega^2_{0}}+O(\lambda^2)
\end{equation}
To understand the structure of  the perturbation theory in terms of diagrams (see Fig 1) ,we denote the lowest order dynamical variable $x_0(\omega)$ by a solid line and the stochastic variable $f(\omega)$ by a wavy line. The lowest order response function $   R(\omega_{0})=\frac{< x(\omega_{0})f(\omega^{'}) >}{< f(\omega)f(\omega^{'}) >}$ is denoted by a half straight and half wavy line and the full response function $R(\omega)$ by a half wavy and a half solid double line. The lowest order correlation function $   C(\omega_{0})=\frac{< x_0(\omega)x_0(\omega^{'}) >}{< f(\omega)f(\omega^{'}) >}$ will be denoted by a solid line with a circle. The action of nonlinearity on the dynamics $(\ddot{x}=-\lambda x^3)$ will be denoted by a dark circle from which three $x_0(\omega)$ lines emanate (or meet) and the fourth line is a lowest order response function $R_0(\omega)$.The dictionary is shown in Fig 1.
\begin{center}
\includegraphics[width=1.1\linewidth]{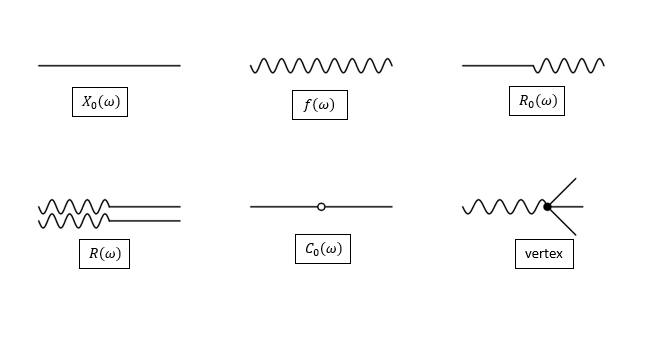}
\captionof{figure}{Dictionary Feynman diagrams}
\end{center}
We now show what the first order perturbation theory result looks like diagrammatically using the above rules. To $O(\lambda)$ ,we have ( using Eq.\ref{eq:19} and Eq.\ref{eq:21} )
\begin{align}
\label{eq:24}
    R(\omega)&=R_{0}(\omega)+\lambda R_1(\omega)+O(\lambda^2) \nonumber\\
    &=R_{0}(\omega)-3\lambda R_{0}(\omega)\frac{D}{2\Gamma \omega^2_0}R_{0}(\omega)+O(\lambda^2)
\end{align}
The second term on the right hand side of Eq.\ref{eq:24} has three distinct features relative to the first term. We begin with a $x_0(\omega)$ line, which through a $R_0(\omega)$ gives rise to an interaction vertex where three solid $x_0$ lines are produced having the frequencies $\omega_1$,$\omega_2$ and $(\omega-\omega_1-\omega_2)$. One of the $x_0$ lines ( say the one with frequency $(\omega-\omega_1-\omega_2)$) joins with the $f(\omega^{'})$ giving another $R_0$ and the other two join with each other to give rise to a correlation function $C(\omega_1)$  with the frequency integrated over. The process is shown in Fig.2.
\begin{center}
\includegraphics[width=1.1\linewidth]{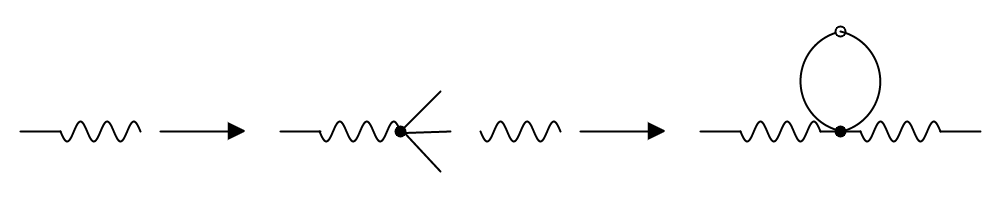}
\captionof{figure}{}
\end{center}
  The factor of 3 comes from the fact that the stochastic term $f(\omega^{'})$ could have coupled with any one of the three $x_0(\omega)$ ’s. The result shown in Eq.\ref{eq:23} have been obtained by various techniques before. The efficacy of the diagrammatic perturbation theory shown here can be seen when we consider the second order perturbation theory and find in addition to a further contribution to the frequency, there is an additional  contribution to the damping and further the correction to the oscillation frequency is dependent on the external frequency $\omega$.This correction to the damping has not been calculated before but anticipated in the study of the effect of noise on the hysteresis in a periodically forced stochastic oscillator \cite{S K Ma}. The fact that the responses can depend upon the probing frequency is a consequence of the systematic perturbation theory.\\
  At $O(\lambda^2)$ ,we need to calculate $R_2(\omega)$ which from Eq.\ref{eq:14} can be written as
  \begin{equation}
       R_2(\omega)=\frac{<x_2(\omega)f(\omega^{'})>}{<f(\omega)f(\omega^{'})>}
  \end{equation}
  The expression for $x_2(\omega)$ can be read off from Eq.\ref{eq:15.c} as 
\begin{equation}
    x_2(\omega)=-3R_0(\omega)\int\frac{d\omega_1}{2\pi}\frac{d\omega_2}{2\pi}x_0(\omega_1)x_0(\omega_2)x_1(\omega-\omega_1-\omega_2)
\end{equation}
Using  the expression for $x_1(\omega)$ from Eq.\ref{eq:15.c}, we obtain
\begin{equation}
\label{eq:27}
    x_2(\omega)=3R_0(\omega)\int\frac{d\omega_1}{2\pi}\frac{d\omega_2}{2\pi}\frac{d\omega^{'}_1}{2\pi}\frac{d\omega^{'}_2}{2\pi}x_0(\omega_1)x_0(\omega_2)R_{0}(w)x_0(\omega^{'}_1)x_0(\omega^{'}_2)x_0(w-\omega^{'}_1-\omega^{'}_2)
\end{equation}
where the frequency $w$ stands for $w=\omega-\omega_1-\omega_2$. The above equation shows that in the calculation of $R_2(\omega)$,we need the correlation
function\\ $<x_0(\omega_1)x_0(\omega_2)x_0(\omega^{'}_1)x_0(\omega^{'}_2)x_0(\omega-\omega_1-\omega_2-\omega^{'}_1-\omega^{'}_2)f(\omega^{'})>$ multiplied by the response function $R_0(\omega-\omega_1-\omega_2)$.This can be done in two different ways:\\
 \raggedright {\textbf{A) Irreducible diagrams:}}\\
  In this category there are two types of diagrams
\renewcommand{\theenumi}{\Roman{enumi}}%
\begin{enumerate}
  \item[i)] If noise term $f(\omega^{'})$ groups with anyone of the three $x_0(\omega^{'}_1)$,  $x_0(\omega^{'}_2)$ and $x_0(\omega-\omega_1-\omega_2-\omega^{'}_1-\omega^{'}_2)$.Remaining two of the three will group with $x_0(\omega_1)$ and $x_0(\omega_2)$ respectively. We get Figure.3a. 
  \item[ii)] Another possible diagram is, if the noise term $f(\omega^{'})$ groups with anyone among $x_0(\omega_1)$,$x_0(\omega_2)$. And the remaining $x_0$ groups with one of the three $x_0$'s of the next vertex (i.e  $x_0(\omega^{'}_1)$,  $x_0(\omega^{'}_2)$ or $x_0(\omega-\omega_1-\omega_2-\omega^{'}_1-\omega^{'}_2)$) forming a correlator and other remaining two $x_0$'s of the same vertex group among themselves forming another correlator.(See Figure.3b)
\end{enumerate}
\begin{center}
\includegraphics[width=0.9\linewidth]{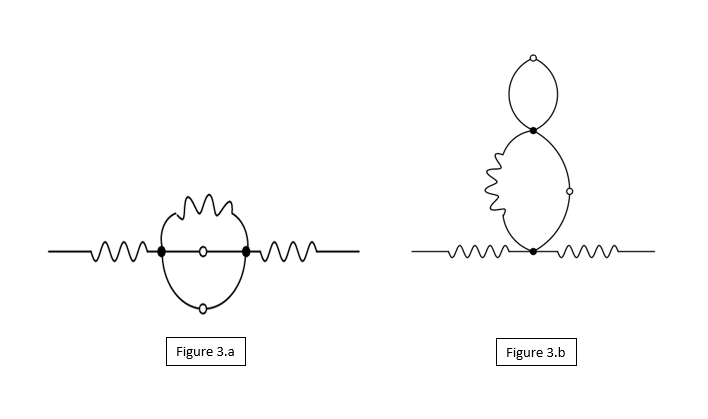}
\end{center}
 \raggedright {\textbf{B)	Reducible diagrams:}}\\
 These diagrams can be cut into two by drawing a line across a connector. In the second order case, as shown in Fig.(3c), it is the square of the first order contribution . This is in accordance with what was noted in Sec.I . These diagrams which repeat a basic block appear at all orders and constitute a geometric series for the block. It allows for a resummation of the series establishing  Dyson’s equation which ensures that at every order of perturbation theory, the block which appears with increasing powers in the subsequent orders is the unit which provides the correction to the zeroth order frequency and damping at that order. We can write the inverse propagator as\\
 $R^{-1}=R_0^{-1}+\Sigma(\omega)=\omega^2_0-\omega^2+2i\Gamma\omega+\Sigma(\omega)$.We actually calculate $R(\omega)$ in perturbation theory and then do the inversion. The irreducible part of the perturbation theory expressions  constitute the $\Sigma(\omega)$ - the real part gives the correction to the frequency and the  imaginary part the correction to the damping coefficient.
 \begin{center}
\includegraphics[width=0.7\linewidth]{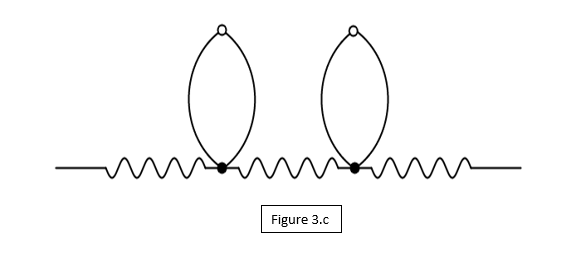}
\end{center}
Returning to the second order diagram (Fig 3.a), we note that the lines with the  two frequencies $\omega_1$,$\omega_2$  at the left vertex can combine with two of the three lines at the right vertex in 6 ways and the remaining line couples to $f(\omega^{'})$. Thus the correlation function becomes
\begin{align}
\label{eq:28}
  <x_0(\omega_1)x_0(\omega_2)x_0(\omega^{'}_1)x_0(\omega^{'}_2)x_0(\omega-\omega_1-\omega_2-\omega^{'}_1-\omega^{'}_2)f(\omega^{'})> \nonumber&=\\6<x_0(\omega_1)x_0(\omega^{'}_1)><x_0(\omega_2)x_0(\omega^{'}_2)><x_0(\omega-\omega_1-\omega_2-\omega^{'}_1-\omega^{'}_2)f(\omega^{'})>  \nonumber&=\\
6C_{0}(\omega_1)C_{0}(\omega_2)R_0(\omega)<f(\omega)f(\omega^{'})>
\end{align}
To obtain the correlation function $R_2(\omega)$,we need to multiply the right hand side of Eq.\ref{eq:28} by $R_0(\omega-\omega_1-\omega_2)$ and divide by $<f(\omega)f(\omega^{'})>$. (see Eq.\ref{eq:27}) This algebra immediately leads to
\begin{align}
\label{eq:29}
     R_2(\omega)&=\frac{18}{(\omega^2_0-\omega^2+2i\Gamma\omega)^2}\int\frac{d\omega_1}{2\pi}\frac{d\omega_2}{2\pi}C_0(\omega_1)C_0(\omega_2)R_0(\omega-\omega_1-\omega_2)\nonumber\\
    &=18R^2_0(\omega)\int_{-\infty}^{\infty}\frac{d\omega_1}{2\pi}C_0(\omega_1)I(\omega_1)
\end{align}
The integral $I(\omega_1)$ is
\begin{equation}
\label{eq:30}
    I(\omega_1)=\int \frac{d\omega_2}{2\pi}C_0(\omega_2)R_0(\omega-\omega_1-\omega_2)
\end{equation}
In the above integrand, we write $\Omega=\omega-\omega_1$  and we get
\begin{align}
\label{eq:31}
I(\omega_1)=\frac{-2D}{8\Gamma\omega^2_0}\frac{\pi(\Omega^2+8\Gamma^2)(\Omega^2-4\omega^2_0)+8\pi^2\Gamma^2\Omega^2+2i\Gamma\Omega(\Omega^2+4\omega^2_0+16\Gamma^2)}{[\Omega^2+4\Gamma^2][(\Omega^2-4\omega^2_0)^2+16\Gamma^2\Omega^2]}
\end{align}
The new feature of the above second order result is that it gives a correction to the damping coefficient $\Gamma$ . Comparing with Eq.\ref{eq:19}, we note that the correction $\Gamma_2$ to the linear order damping is given by
\begin{equation}
    \Gamma_2=\frac{9D^2}{\omega}\int_{-\infty}^{\infty}\frac{d\omega_1}{2\pi\omega^2_0}\frac{\Omega(\Omega^2+4\omega^2_0+16\Gamma^2)}{[\Omega^2+4\Gamma^2][(\Omega^2-4\omega^2_0)^2+16\Gamma^2\Omega^2][(\omega^2_0-(\Omega-\omega)^2)^2+4(\Omega-\omega)^2\Gamma^2]}
\end{equation}
We can express the above equation as:
\begin{equation}
\label{eq:33}
    \Delta\Gamma=\lambda^2\Gamma_2=\frac{9\lambda^2D^2}{2\pi\omega}J(\omega)
\end{equation}
   The  integral $J(\omega)$ spans half the real axis and is given by
   \begin{subequations}
   \begin{equation}
   \label{eq:34.a}
       J(\omega)=\frac{8\omega}{\omega^2_0}\int_{0}^{\infty}d\Omega\frac{\Omega^2(\Omega^2+4\omega^2_0+16\Gamma^2)(2\Gamma^2+\Omega^2+\omega^2-\omega^2_0)}{(\Omega^2+4\Gamma^2)[(\Omega^2-4\omega^2_0)^2+16\Gamma^2\Omega^2]F(\Omega)F(-\Omega)}
   \end{equation}
   The function $F(\Omega)$ is found to be
   \begin{equation}
   \label{eq:34.b}
       F(\Omega)=[(\omega^2_0-(\omega-\Omega)^2)^2+4\Gamma^2(\omega-\Omega)^2]
   \end{equation}
   \end{subequations}
   
   The integral in Eq.\ref{eq:34.a} can be approximately evaluated by noting that for low damping ($\Gamma<<\omega_0$ ), the contribution near $\Omega=2\omega_0$ is the maximum and hence we find
   \begin{equation}
   \label{eq:25}
       J(\omega)=\frac{2\pi\omega(3\omega^2_0+\omega^2)}{\Gamma\omega^2_0(9\omega^4_0+\omega^4-10\omega^2\omega^2_0)^2}
   \end{equation}
   We find that at low frequencies ($\omega << \omega_0$),$J(\omega) \approx \frac{2\pi\omega}{81\Gamma\omega^8_0}$ and at high frequencies ($\omega >> \omega_0$), $J(\omega) \approx \frac{2\pi}{\Gamma\omega^2_0\omega^5}$ . Thus we get the renormalized damping coefficient $\Gamma_{R}$ for the cubic stochastic oscillator to be given by the following expressions:  
   \[
     \Gamma_{R}\approx     
     \begin{dcases}
     \Gamma+\frac{\lambda^2D^2}{9\Gamma\omega^8_0} ,&\omega <<\omega_0 \\
     \Gamma+\frac{9\lambda^2D^2}{\Gamma\omega^2_0\omega^6} ,&\omega >>\omega_0
     \end{dcases}
\]
The fact that the damping coefficient is renormalized at O($\lambda^2$) and that the renormalization is frequency dependent is one of the most significant results of the diagrammatic technique.\\
For completeness, we also  write down the two loop correction $\omega^2_2$ to the frequency of oscillation arising from Fig.(3a). This is given by the expression 
\begin{equation}
    \Delta_1=9D^2\lambda^2J^{'}(\omega)
\end{equation}
The integral $J^{'}(\omega)$ is given by the expression
\begin{equation}
    \int_{-\infty}^{\infty}\frac{d\Omega}{2\Gamma\omega^2_0}\frac{(\Omega^2+8\Gamma^2)(\Omega^2-4\omega^2_0)+8\pi\Gamma^2\Omega^2}{[\Omega^2+4\Gamma^2][(\Omega^2-4\omega^2_0)^2+16\Gamma^2\Omega^2][(\omega^2_0-(\Omega-\omega)^2)^2+4(\Omega-\omega)^2\Gamma^2]}
\end{equation}
Once again noting that for $\Gamma << \omega_0$ , the primary contribution to the integral comes from the region around $\Omega=2\omega_0$ ,we find that 
\begin{equation}
    J^{'}(\omega)=\frac{\pi^2}{8\omega^4_0(3\omega^2_0-4\omega\omega_0+\omega^2)^2}
\end{equation}
We now return to the diagram of Fig.(3b) and note that unlike the double frequency integration of the diagram of Fig.(3a), here we have a product of two integrals. The contribution $\Delta_2$ of this diagram to $R(\omega)$ can be read off from Fig.(3b) as 
\begin{align}
    \Delta_2&=36\lambda^2 D^2 R^2_0(\omega)\int\frac{d\omega_1}{2\pi}\frac{1}{(\omega^2_1-\omega^2_0)^2+4\Gamma^2\omega^2_1}\int\frac{d\omega_2}{2\pi}\frac{(\omega^2_0-\omega^2_2)-2i\lambda\omega_2}{((\omega^2_2-\omega^2_0)^2+4\Gamma^2\omega^2_2)^2} \nonumber \\
    &=\frac{9\lambda^2D^2}{4\omega^6_0\Gamma^2}R^2_0(\omega)
\end{align}
Clearly, this diagram gives no contribution to the damping but gives a negative contribution to the frequency of oscillation with the total contribution $\omega^2_2$ given by 
\begin{equation}
    \omega^2_2=9D^2(J^{'}(\omega)-\frac{1}{4\omega^6_0\Gamma^2})
\end{equation}
We conclude this section with an intriguing observation. It is apparent by looking at the function within the brackets in the equation above, that at all frequencies the second term dominates. This allows an approximation where at every order of perturbation theory we keep only the diagrams which are essentially the repeat of Fig. 3b.  In Fig.4 we show a sequence of diagrams of the kind shown in Fig 3b at all orders of perturbation theory which can be summed since they follow a geometric progression.
\begin{center}
\includegraphics[width=0.9\linewidth]{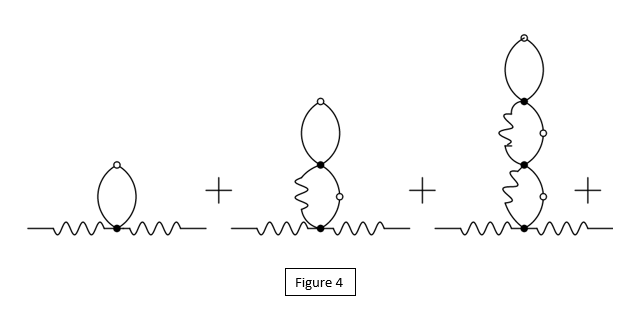}
\end{center}
 If we consider only this infinite set of diagrams then they lead to the effective final frequency $\Omega$ given by the equation 
 \begin{align}
 \label{eq:41}
     \Omega^2&=\omega^2_0+\frac{3\lambda D}{2\Gamma \omega^2_0}-\frac{9\lambda^2D^2}{4\Gamma^2\omega^6_0}+\frac{27\lambda^3D^3}{8\Gamma^3\omega^{10}_0}-.....\nonumber\\
     \Delta\omega^2&=\frac{3\lambda D}{2\Gamma\omega^2_0}\Big[ 1-\frac{3\lambda D}{2\Gamma \omega^4_0+...}\Big]=\frac{3\lambda D}{2\Gamma\omega^2_0}\frac{1}{(1+\frac{3\lambda D}{2\Gamma \omega^4_0})}
 \end{align}
 This is to be contrasted with the non-stochastic oscillator with cubic non-linearity where the frequency increases as  $(\lambda E)^{\frac{1}{4}}$ for large $\lambda E$, where E is the energy of oscillation.
 
 The somewhat intriguing answer of Eq.\ref{eq:41} can be found in another context. This is the spherical limit [25-27] which was popular in the study of critical phenomena and in the early days of lattice gauge theories. In our case, we can study this limit by considering the dynamics in a N-dimensional space where our scalar variable $x$ becomes a N-component vector $\{x_i\},\, i=1,2,...N$  and the dynamics is given by     
 \begin{equation}
     \ddot{x_i}+2\Gamma\dot{x_i}+\omega^2_0x_i=-\lambda\frac{x_j x_j x_i}{N}+f_i
 \end{equation}
The noise $f_i$ is a N-component vector and and we have used the repeated index summation convention which implies that a repeated index has to interpreted as a sum from 1 to N. Note that the coupling constant is now $\lambda/N$ in order to compensate for the fact that $x_j x_j$ has N terms and the noise correlation carries a factor $\delta_{ij}$ indicating that only the noise terms with same subscript have nonzero correlation. We consider the response function $R_i(\omega)$ for a fixed $i$ which let us say is $i=1$ . At the zeroth order it is simply 
	\begin{equation}
	    R_{01}=\frac{1}{(\omega^2_0-\omega^2+2i\Gamma\omega)}
	\end{equation}                         
If we consider the one loop graph of Fig 2 then there are N such diagrams corresponding to the internal line having subscripts going from 1 to N and hence the factor of $N^{-1}$ is cancelled. Thus this diagram survives if $N\to \infty$ .We now look at the two loop diagrams of Fig 3a and Fig 3b.Considering Fig.3a, in the loop the correlators have to carry the same subscript and hence there are N diagrams . The coupling constant occurs twice and is of order $N^{-2}$. This diagram does not survive in the  $N\to \infty$ limit ( spherical limit ).Now, we consider Fig 3b and the both the bottom and top loops carry a multiplicative factor of N . The two coupling constants yield a factor of $N^{-2}$ and this diagram is of $O(1)$. The trend should be clear- in the limit of $N\to \infty$, the only survivors are the set of diagrams shown in Fig.4 and we have  the result of Eq.\ref{eq:41} once again!Clearly in this limit the damping is zero , the first non-vanishing correction to the damping in this framework is  $O(N^{-1})$- reminiscent of critical phenomena . The damping correction is the direct analogue of the anomalous dimension $\eta$ in critical phenomena [19].
 \section{ The stochastic oscillator with periodic forcing}
 We begin by clarifying the terminology – the equation of motion written down in Eq.\ref{eq:1} defines the stochastic oscillator with cubic nonlinearity. The Duffing oscillator, in the terminology followed here, is the oscillator with cubic nonlinearity  forced by an external oscillatory force $A cos({\Omega t})$  where A is the constant amplitude and $\Omega$ is the forcing frequency[28-31]. Accordingly our stochastic Duffing oscillator has the equation of motion
 \begin{equation}
\label{eq:42}
    \ddot{x}+2\Gamma\dot{x}+\omega_0^2x+\lambda x^3=f(t)+F(t)
\end{equation}
with the external force $F(t)$ prescribed as 
\begin{equation}
    F(t)=Acos({\Omega t})=\frac{A}{2}(e^{i\Omega t}+e^{-i\Omega t})
\end{equation}
In frequency-space, Eq.\ref{eq:42} now becomes
\begin{align}
    \Big [ (\omega^2_0-\omega^2)+2i\Gamma \omega \Big] x(\omega)= \frac{A}{2}\Big[ \delta(\omega-\Omega)+\delta(\omega-\Omega)\Big]+f(\omega)\nonumber\\-\lambda\int\frac{d\omega_1}{2\pi}\frac{d\omega_2}{2\pi}x(\omega_1)x(\omega_2)x(\omega-\omega_1-\omega_2)
\end{align}
    It should be noted that the zeroth order response function $R_0(\omega)$  remains the same as before.\\
The usual perturbation series for $x(\omega)$  can be written down as
\begin{equation}
\label{eq:45}
    x(t)=x_0(t)+\lambda x_{1}(t)+\lambda^2 x_2(t)+....
\end{equation}
The zeroth order term is seen to be 
\begin{equation}
\label{eq:46}
    x_0(\omega)=\frac{A}{2}\Big[\frac{\delta(\omega+\Omega)}{\omega^2_0-\Omega^2-2i\Gamma\Omega}+\frac{\delta(\omega-\Omega)}{\omega^2_0-\Omega^2+2i\Gamma\Omega}\Big]+\frac{f(\omega)}{\omega^2_0-\omega^2+2i\omega\Gamma}
\end{equation}
The first order contribution  to the response function is obtained, as before,from $R_1(\omega)=\frac{<x_1(\omega)f(\omega^{'})>}{<f(\omega)f(\omega^{'})>}$  with the term $x_1(\omega)$ obtained from the integral
\begin{equation}
\label{eq:47}
     x_1(\omega)=-R_0(\omega)\int\frac{d\omega_1}{2\pi}\frac{d\omega_2}{2\pi}x_{0}(\omega_1)x_{0}(\omega_2)x_{0}(\omega-\omega_1-\omega_2)
\end{equation}
The difference from the first order response calculated in Sec II lies in  the extra term proportional to $A$ which we have in Eq.\ref{eq:46}. This implies that when we calculate $R_1(\omega)$ we will have the term shown in Eq.\ref{eq:21} and in addition pick up terms which vanish if we set $A=0$. Explicitly,
\begin{align}
    <x_1(\omega)f(\omega^{'})>&=-3R_0(\omega)\int\frac{d\omega_1}{2\pi}\frac{d\omega_2}{2\pi}<x_0(\omega_1)x_0(\omega_2)><x_0(\omega-\omega_1-\omega_2)f(\omega^{'})>\nonumber\\
    &=-3R_0(\omega)\int\frac{d\omega_1}{2\pi}\frac{d\omega_2}{2\pi}<x_0(\omega_1)x_0(\omega_2)>\delta(\omega_1+\omega_2)R_0(\omega)<f(\omega)f(\omega^{'})>
\end{align}
Using Eq.\ref{eq:46}, we see that there are two kinds of non-vanishing terms ( no $f(\omega)$ or two $f(\omega^{'})$ ’s in the expectation value $<x_0(\omega_1)x_0(\omega_2)>$ . The term with two $f(\omega)$'s has already been treated in Eq.\ref{eq:16b}. Here we focus on terms with no $f(\omega)$'s. Since the $x_0$-correlator survives only for $\omega_1=-\omega_2$,we see that in the product of the two $x_0$'s, the term with the coefficient $\frac{A^2}{4}$ is the only survivor. This new contribution to $R_1(\omega)$ can be written as 
\begin{equation}
    \frac{A^2}{4}\Big[\frac{\delta(\omega_1+\Omega)}{\omega^2_0-\Omega^2-2i\Gamma\Omega}+\frac{\delta(\omega_1-\Omega)}{\omega^2_0-\Omega^2+2i\Gamma\Omega}\Big]\Big[\frac{\delta(\omega_2+\Omega)}{\omega^2_0-\Omega^2-2i\Gamma\Omega}+\frac{\delta(\omega_2-\Omega)}{\omega^2_0-\Omega^2+2i\Gamma\Omega}\Big]
\end{equation}
and with constraint $\omega_1=-\omega_2$, we have the new correlator $C_{F}(\omega)$ given by
\begin{equation}
\label{eq:50}
    C_{F}(\omega)=<x_0(\omega)x_0(\omega^{'})>=\frac{2D\delta(\omega+\omega^{'})}{(\omega^2_0-\omega^2)^2+4\Gamma^2\omega^2}+\frac{A^2}{2}\frac{\delta(\omega+\omega^{'})}{(\omega^2_0-\Omega^2)^2+4\Gamma^2\Omega^2}
\end{equation}
Because of the delta function constraints, the survivors are the terms for which $\omega_1=-\omega_2=\pm\Omega$. We thus get an additional contribution to $<x_1(\omega)f(\omega^{'})>$ beyond what we had in Eq.\ref{eq:21} This new contribution can be written  as (using Eq.\ref{eq:47}) $\frac{-3}{2}R^2_0(\omega)A^2[{(\omega^2_0-\Omega^2)^2+4\Omega^2\Gamma^2}]^{-1}$ .The complete $R_1(\omega)$ for the stochastic Duffing oscillator at the one loop order is consequently
\begin{equation}
    R_1(\omega)=\frac{-3R^2_0D}{2\Gamma\omega^2_0}-\frac{3}{2}R^2_0(\omega)A^2\frac{1}{(\omega^2_0-\Omega^2)^2+4\Omega^2\Gamma^2}
\end{equation}
The one-loop frequency for the harmonically driven nonlinear oscillator ( cubic non-linearity) is consequently given by $\omega^2=\omega^2_0+\lambda\omega^2_1$ where
\begin{equation}
\label{eq:52}
    \omega^2_1=\frac{3D}{2\Gamma\omega^2_0}+\frac{3A^2}{2[(\omega^2_0-\Omega^2)^2+4\Gamma^2\Omega^2]}
\end{equation}
The  additive  contributions of the regular and stochastic forcing is a special feature  of the one loop result. At the two-loop level when we consider the contribution from Fig.(3a), there will be a contribution proportional to $DA^2$ ,which is apparent  from the structure of the product $R_0(\omega-\omega_1-\omega_2)C_{F}(\omega_1)C_{F}(\omega_2)$ . The calculation is involved for this diagram but straightforward for the diagram shown in Fig. (3b). The integral here splits into two separate integrals. The top loop gives the contribution shown in Eq.\ref{eq:52}, while the bottom loop is the integral $\int d\omega C_{F}(\omega)R_0(-\omega)$ . The correlator $C_{F}(\omega)$ is given in Eq.\ref{eq:50} and the propagator is the usual $R_0(\omega)$ . Consequently the integral has the same structure as shown in Eq.\ref{eq:52}. The product of two factors which have the structure shown in Eq.\ref{eq:52} leads to the term of the form $DA^2$ .This is a feature which this systematic perturbation theory has revealed. 

\section{Escape due to stochasticity}
In this section, we consider a situation which is generally the domain of non-equilibrium statistical mechanics – a situation where a particle can overcome a potential barrier at a finite temperature $T$ because of the fluctuations induced by the temperature. Higher the temperature, higher is the ability to overcome the barrier. The situation is analysed in terms of the diffusion coefficient $\Bar{D}$  which is proportional to the strength $D$ of the fluctuations that we have considered here and the relaxation rate $\Gamma$ and leads to the extremely striking result $\frac{\Bar{D}}{\Gamma}=k_B T$ which is known as the fluctuation dissipation theorem. We will look at this from the dynamics of a particle trapped in the well of a cubic potential (shown in Fig.5) and subjected to a stochastic forcing. The point that we want to make is that even if the average value of the displacement of the particle is less than that corresponds to the peak of the potential, the fluctuation about the mean can take the particle beyond the maximum of the potential. The cubic potential is taken to be
\begin{equation}
    V(x)=\frac{m\omega^2_0x^2}{2}-\frac{m\mu x^3}{3}
\end{equation}
\begin{center}
\includegraphics[width=0.9\linewidth]{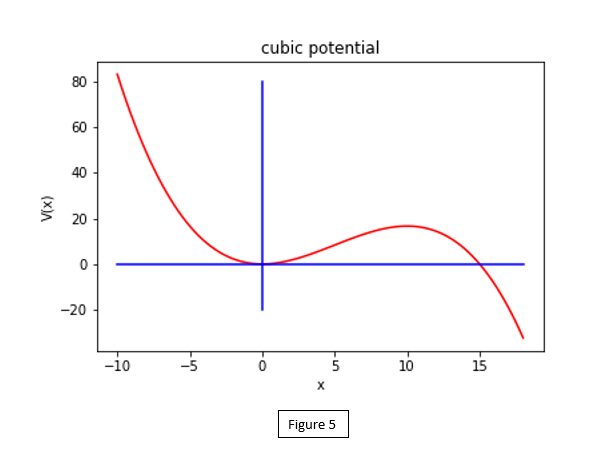}
\end{center}
and the initial position of the particle is within the well.  In the presence of a stochastic force the equation of motion is 
\begin{equation}
    \ddot{x}+2\Gamma\dot{x}+\omega^2_{0}x-\mu x^2=f(\omega)
\end{equation}
As before we expand 
\begin{equation}
\label{eq:55}
    x=x_0+\mu x_1+\mu^2 x_2+....
\end{equation}
In the frequency space $x_0=f(\omega)[\omega^2_0-\omega^2+2i\omega\Gamma]^{-1}$ at the zeroth order. At
 the first order in $\mu$,we get
\begin{align}
    x_1(\omega)&=R_0(\omega)\int_{-\infty}^{\infty}\frac{d\omega_1}{2\pi}x_0(\omega_1)x_0(\omega-\omega_1)\nonumber\\
    &=R_0(\omega)\int_{-\infty}^{\infty}\frac{d\omega_1}{2\pi}\frac{f(\omega_1)f(\omega-\omega_1)}{[\omega^2_0-\omega^2+2i\Gamma\omega_1][\omega^2_0-(\omega-\omega_1)^2+2i\Gamma(\omega-\omega_1)]}
\end{align}
The average value of the displacement at this order is
\begin{align}
\label{eq:57}
    <x_1(\omega)>&=2DR_0(\omega)\delta(\omega)\int\frac{d\omega_1}{2\pi}\frac{1}{(\omega^2_0-\omega^2_1)^2+4\Gamma^2\omega^2_1}\nonumber\\
    &=\frac{D}{2\Gamma\omega^4_0}
\end{align}
Since the average position of the particle is at $O(\mu)$,it is likely to be far from the position of the barrier which is at $x_0=\omega^2/\mu$  and should not be able to escape from the well. However, there is a fluctuation round the mean which can be quite large. The mean square fluctuation is $<(x-<x>)^2>=<x^2>-<x>^2$ and at the lowest order $<x^2>=2DC_0(\omega_1)$ ,while $<x>^2$ is $O(\mu^2)$ .Since the lowest order value of $<x^2>$ is $\frac{D}{2\Gamma\omega^2_0}$ , the magnitude $\Delta_1$ of the fluctuation is 
\begin{equation}
    \Delta=\sqrt{\frac{D}{2\Gamma\omega^2_0}}+O(\mu^2)
\end{equation}
The ratio of the fluctuation to the mean is given by 
\begin{equation}
\label{eq:59}
    \frac{\Delta}{<x>}=\frac{\sqrt{2}\omega^3_0}{\mu}\sqrt{\frac{\Gamma}{D}}
\end{equation}
To escape from the well the maximum potential barrier  $V_0$ that the particle has to overcome is $V_0=\frac{\omega^6}{6\mu^2}$  which makes the right hand side of Eq.\ref{eq:59} have the form $2\sqrt{\frac{3V_0\Gamma}{D}}$.Clearly, for the particle to have the ability to escape, the quantity $<x>+\Delta$ has to exceed $\frac{\omega^2_0}{\mu}$ . This makes the condition 
\begin{equation}
    \mu<x_1>+\Delta=\frac{\mu D}{2\Gamma\omega^4_0}+\sqrt{\frac{D}{2\Gamma\omega^2_0}}>\frac{\omega^2_0}{\mu}
\end{equation}
a very good measure of the ability to escape.Similar kind of escape condition can also be found in quantum case \cite{A Pal}. \\
We end this section by demonstrating how the perturbation theory works in this case. We wrote down  $x_1(\omega)$  in Eq.\ref{eq:57}. Here we write down the two succeeding orders as 
\begin{align}
    x_2(\omega)&=2R_0(\omega)\int\frac{d\omega_1}{2\pi}x_0(\omega_1)x_1(\omega-\omega_1)\nonumber\\
    x_3(\omega)&=R_0(\omega)\Big[\int\frac{d\omega_1}{2\pi}x_1(\omega_1)x_1(\omega-\omega_1)+2\int\frac{d\omega_1}{2\pi}x_0(\omega-\omega_1)x_2(\omega_1)\Big]
\end{align}
We can write the mean square displacement as: \begin{equation}
    <x^2>=\int\frac{d\omega}{2\pi}<x(\omega)x(-\omega)>=\frac{D}{2\Gamma\omega^2_0}+\mu^2\int\frac{d\omega}{2\pi}\Big[<x_1(\omega)x_1(-\omega)>+2<x_2(\omega)x_0(-\omega)>\Big]
\end{equation}
For the mean displacement $<x>=\mu<x_1>+\mu^3<x_3>+...$ we can use the expansion of Eq.\ref{eq:55} to write
\begin{equation}
\label{eq:63}
    <x_3>=\int\frac{d\omega}{2\pi}<x_3(\omega)>=\int\frac{d\omega^{'}}{2\pi}\frac{d\omega}{2\pi}R_0(\omega)[<x_1(\omega^{'})x_1(\omega-\omega^{'}))>+2<x_2(\omega^{'})x_0(\omega-\omega^{'}))>]
\end{equation}
The diagrammatic representations of the two terms on the right hand side of $<x_3(\omega)>$ are as shown in Fig. 6a  and 6b. 
\begin{center}
\includegraphics[width=0.9\linewidth]{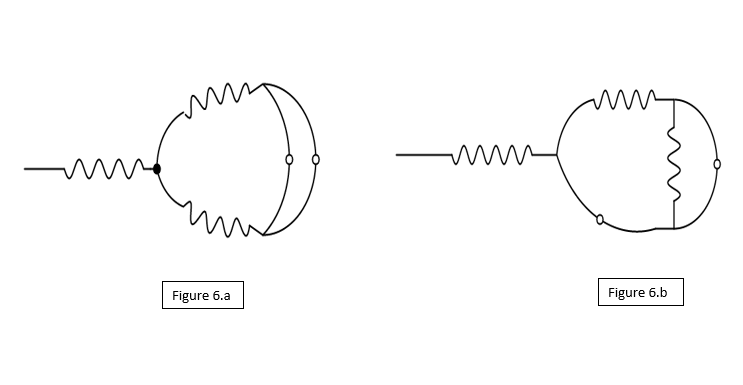}
\end{center}
Computing the first term of Eq.\ref{eq:63}.
\begin{align}
    <x_1(\omega^{'})x_1(\omega-\omega^{'})>&=\int\frac{d\omega_2}{2\pi}\frac{d\omega_1}{2\pi}R_0(\omega^{'})R_0(\omega-\omega^{'})<x_0(\omega_1)x_0(\omega-\omega^{'})x_0(\omega_2)x_0(\omega-\omega^{'}-\omega_2)>\nonumber\\
    &=2R_0(\omega^{'})R_0(\omega-\omega^{'})\int\frac{d\omega_1}{2\pi}C_0(\omega_1)C_0(\omega^{'}-\omega_1)\delta(\omega)
\end{align}
As for the second expectation value we have, using Fig.6b
\begin{equation}
    <x_2(\omega^{'})x_0(\omega-\omega^{'})>=4R_0(\omega^{'})C_0(\omega^{'})\int\frac{d\omega_2}{2\pi}R_0(\omega_2)C_0(\omega_2)\delta(\omega)
\end{equation}
The expectation value $<x_3>$ is found as 
\begin{equation}
    <x_3>=R_0(0)\Big[\int\frac{d\omega^{'}}{2\pi}R_0(\omega^{'})R_0(\omega-\omega^{'})\int \frac{d\omega_1}{2\pi}C_0(\omega_1)C_0(\omega^{'}-\omega_1)+8(\int\frac{d\omega}{2\pi}C_0(\omega_2)R_0(\omega_2))^2\Big]
\end{equation}
An evaluation of the integrals in the small damping limit yields $<x_3>=\frac{D^3}{256\omega^6_0\Gamma^4}$  and hence the average displacement is
\begin{equation}
    <x>=\mu\frac{D}{2\Gamma\omega^4}+\mu^3\frac{D^3}{256\omega^6_0\Gamma^4}+....
\end{equation}
What is interesting about the result is that expansion is in powers of the dimensionless number $\frac{\mu^2D^2}{\omega^2_0\Gamma^3}$ .For a strongly damped system with high natural frequency, the higher order corrections are strongly suppressed. Hence the criterion for escape from the well is very well represented by the lowest order result ( noting that the well depth is $\frac{\omega^6_0}{6\mu^2}$  which we will denote as $V_0$ )
\begin{equation}
    \frac{D}{\Gamma V_0}\geq 7.2
\end{equation}
Since $V_0$ has the dimensions of energy, so must the combination $\frac{D}{\Gamma}$. This energy is causing the particle to cross a barrier and hence it is tempting to relate it to the temperature T of the system. Hence this little analysis of stochastic dynamics hints at an interesting statistical mechanical result-the fluctuation dissipation theorem.
\section{The  pendulum with random vibration of the point of support}
In this section we consider a variation on the Kapitza pendulum which is a pendulum pointing vertically upwards from its point of suspension with the point of suspension being oscillated vertically . For frequencies higher than a critical value , the upside-down pendulum can execute small oscillations about its mean position. Extensions to the case where the point of support is subjected to random vibration was considered by Shapiro and Loginov \cite{V E Shapiro}and Landa and Zaikin\cite{P S Landa}. It was shown that for random vertical vibrations of the point of support, this pendulum can execute oscillations about the inverted position when the root mean square vertical oscillation vibrates randomly with  larger than a critical amplitude but only if the random vibrations are not $\delta-$correlated.In this section we show how the perturbation theory fits into this scheme of things. The pendulum that we will be working with is described by the equation of motion
\begin{equation}
    \ddot{\theta}+\omega^2_0(1+\epsilon f(t))sin\theta=0
\end{equation}
The angle $\theta$ is our dynamical variable with $\theta=0$ describing the resting position of the normal pendulum and $\theta=\pi$  that of the upside down pendulum.  As can be seen easily ( expanding $sin\theta$ ) the fixed point $\theta=0$ is stable and  $\theta=\pi$ is unstable. We will expand the angle $\theta$  in powers of  $\epsilon$ ( meaning essentially in powers of stochasticity) as 
\begin{subequations}
\begin{equation}
\label{eq:70a}
    \ddot{\theta_0}+\omega^2_0 sin\theta_0=0
\end{equation}
\begin{equation}
\label{eq:70b}
    \ddot{\theta_1}+\omega^2_0(cos\theta_0)\theta_1=-\omega^2_0 sin(\theta_0)f(t)
\end{equation}
\begin{equation}
\label{eq:70c}
    \ddot{\theta_2}+\omega^2_0(cos\theta_0)\theta_2-\frac{\omega^2_1}{2}sin\theta_0=-\omega^2_0 f(t)(cos\theta_0)\theta_1
\end{equation}
\end{subequations}
The average value of $\theta_1$  is clearly zero. However, to get $\theta_2$, whose average value is non-zero, we need an answer for $\theta_1$. We need an approximation to do this. The natural frequency of the pendulum is a finite number of order unity and $sin\theta_0$ is at most unity. The time scale involved on the right hand side is extremely small because the noise term f(t) is rapidly varying. Consequently, to find the relevant ( fast varying ) part of $\theta_1$ , we can take $\ddot{\theta_1}$ to be the dominant term on the L.H.S. of Eq.\ref{eq:70b} and write
\begin{equation}
    \theta_1(t)=-\omega^2_0 sin\theta_0(t)\int^{t}_{0}dt^{'}\int^{t^{'}}_{0}f(t_1)dt_1
\end{equation}
Clearly $<\theta_1>=0$ and this term does not contribute directly to the dynamics of $<\theta>$ but does contribute at the next order.The same logic applied to Eq.\ref{eq:70c} allows us to drop the second term on the left hand side and write
\begin{equation}
\label{eq:72}
    <\ddot{\theta_2}>-\omega^2_0 sin\theta_0\frac{<\theta^2_1>}{2}=\frac{\omega^4_0}{2}sin(2\theta)\Big< f(t)\int^{t}_{0}dt^{'}\int^{t^{'}}_{0}dt_1f(t_1)\Big> 
\end{equation}
We see immediately that if we use the Gaussian white noise with $<f(t)f(t')>=2D\delta(t-t^{'})$ ,then there can be no stabilization.If we use high frequency oscillation for $f(t)$ i.e  $f(t)=A cos\Omega t$ where $\Omega>>\omega$ , we obtain the dynamics of 
\begin{align}
    <\theta>&=\theta_0+\epsilon<\theta_1>+\epsilon^2<\theta_2>+....\nonumber\\
    <\ddot{\theta}>&=\ddot{\theta_0}+\epsilon<\ddot{\theta_1}>+\epsilon^2<\ddot{\theta_2}>+....
\end{align}
The first order in  $\epsilon$ always averages out to zero. The second order yields the effective equation of motion for $\theta$  in the form ( correct to $O(\epsilon^2)$ ) 
\begin{subequations}
\begin{equation}
    \ddot{\theta}+\omega^2_0(1-\epsilon^2\frac{<\theta_1^2>}{2})sin\theta=\epsilon^2 C sin(2\theta)
\end{equation}
\begin{equation}
    C=\frac{\omega^4_0}{2}<\int^{t}_{0}dt_1\int^{t_1}_{0}dt_2 f(t)f(t_2)>
\end{equation}
\end{subequations}
It should be remembered that the final average in these time dependent dynamics is an averaging over the observation time.The sign of the constant C determines whether it is possible to stabilize the upside down position ( namely the fixed point $\theta^{*}=\pi$  ) .It is straightforward to check that for the Kapitza situation where $f(t)=A cos\Omega t$ with $\Omega>>\omega_0$ , the usual answer ( $C<0$ ) is obtained. It is also obvious that for a delta correlated Gaussian noise, $C>0$ and there is no possibility of stabilizing the upside down position.  However,if we use a Gaussian noise with a scale in frequency space  as typified by the two point correlation function
\begin{equation}
\label{eq:75}
    <g(\omega_1)g(\omega_2)>=F(\omega_1)\delta(\omega_1+\omega_2)
\end{equation}
we can find explicitly the constraint under which the stabilization of the inverted position obtains.
Returning to Eq.\ref{eq:72}, we write the correlation function as 
\begin{equation}
\label{eq:76}
    \int^{t}_0 dt_1 \int^{t_1}_0 dt_2 <f(t_1)f(t_2)>=\int^{\infty}_{-\infty}\frac{d\omega_1}{2\pi}\int^{\infty}_{-\infty}\frac{d\omega_2}{2\pi}\int^{t}_{0}dt_1\int^{t_1}_0 dt_2 e^{i\omega_1 t+i\omega_2 t_2}<g(\omega_1)g(\omega_2)>
\end{equation}
In the above, the function $g(\omega)$ is the Fourier transform of the function $f(t)$ with the correlation function as specified in Eq.\ref{eq:75}. With the delta function constraint of Eq.\ref{eq:75}, we can perform the time integrations in Eq.\ref{eq:76} to obtain 
\begin{equation}
    \int^{t}_{0}dt_1\int^{t_1}_{0}dt_2 <f(t)f(t)>=-\int^{\infty}_{-\infty}\frac{d\omega_1}{2\pi}\frac{F(\omega_1)}{\omega^2_1}(1-(1-ti\omega_1)e^{i\omega_1t})
\end{equation}
If we now carry out a long time average over the observation time T, the term involving $exp(i\omega_1 t)$  will average out to zero and we will be left with 
\begin{equation}
    C=-\omega^4_0\int^{\infty}_0\frac{d\omega}{2\pi}\frac{F(\omega)}{\omega^2}
\end{equation}
It is clear that the admissible form of $F(\omega)$ will  have to vanish faster than    $\omega$ near the origin and vanish with a characteristic scale at large frequencies. A possible choice is $F(\omega)=\Gamma^{-3}\omega^{2}exp(-\frac{\omega^2}{\Gamma^2})$ . This leads to $C=-\frac{\omega^4_0}{\sqrt{2\pi^3}\Gamma^2}$ and the stabilization condition becomes $\epsilon^2\Gamma^2/\omega^2_0>\sqrt{2\pi^3}$ , which is a relation similar to the one obtained for the Kapitza pendulum.
\section{Conclusion}
We have developed a systematic perturbation theory for the dynamics of a stochastically driven Duffing oscillator having the dynamics  $\ddot{x}+2\Gamma\dot{x}+\omega^2_0 x+\lambda x^3=f(t)$  where f(t) is the stochastic term with a Gaussian distribution and for the most part delta correlated. Our strategy has  been to base the perturbation theory on the Feynman diagrams which allows us to generate terms which are distinctly different from the first order perturbation theory with its various dressed variants. The emergence of these diagrams ( topologically distinct from the dressed one loop diagrams) leads to small effects which are not accessible from the various approximations available in the current literature. These effects have to do with the linear response function of the above system which (in frequency space) for the linear part of the dynamics is $R_0(\omega)$ and is given by the expression $[\omega^2_0-\omega^2+2i\Gamma\omega]^{-1}$ . The literature has always dealt with the changes to the shift of the resonant frequency from $\omega_0$  due to the nonlinear term. Systematizing the perturbation theory shows that the shift occurs in both $\omega_0$  and the damping coefficient $\Gamma$. Further, we find that the shift is frequency dependent. If we consider the periodically driven system i.e. add a term $Acos(\Omega t)$  to the above dynamics then there is a contribution to the shifts where the periodic and oscillatory drives interact.We have also derived the condition for escape from a cubic meta stable well in presence of stochastic forcing. And interestingly we got the hints of fluctuation dissipation theorem.Using this perturbation theory in a slightly different context for the Kapitza pendulum with the periodic frequency modulation replaced by a noise term f(t), we find that in perturbation theory a delta correlated noise does not produce the Kapitza phenomenon ( stabilization of the upside down position)but a coloured noise certainly does.

\end{document}